\documentclass[a4paper,11pt]{article}
\pdfoutput=1 

\usepackage{jinstpub} 
\usepackage{siunitx}
\usepackage{circuitikz}
\usepackage{hyperref}
\usepackage{url}
\usepackage{acro}
\DeclareAcronym{SAMPIC}{
	short = SAMPIC,
	long  = SAMpler for PICosecond
}
\DeclareAcronym{OPTIMA}{
	short = OPTIMA,
	long  = Optimized Precision Timing for Multichannel Acquisition
}
\DeclareAcronym{DUT}{
	short = DUT,
	long  = Device Under Test
}
\DeclareAcronym{ROI}{
	short = ROI,
	long  = Region Of Interest
}
\DeclareAcronym{MCP}{
	short = MCP,
	long  = Microchannel Plate
}
\DeclareAcronym{CFD}{
    short = CFD,
    long  = Constant Fraction Discriminator
}
\notoc%

\usepackage{lineno}

\title{OPTIMA, a board dedicated to Optimized Precision Timing for Multichannel Acquisition}

\author[a,b,1]{Federico De Benedetti,\note{Corresponding author.}}
\author[a]{Victor Coco,}
\author[a]{Paula Collins,}
\author[a]{Raphael Dumps,}
\author[a]{Edgar Lemos Cid,}
\author[a,c]{Alfonso Puicercus Gomez,}
\author[a]{Efr\'en Rodr\'iguez Rodr\'iguez,}
\author[a]{Morag Williams}

\affiliation[a]{CERN,\\Esplanade de Particules 1 Meyrin, Switzerland}
\affiliation[b]{Universidade de Santiago de Compostela,\\Praza do Obradoiro 0, Santiago de Compostela, Spain}
\affiliation[c]{Van Swinderen Institute, University of Groningen,\\Broerstraat 5,Groningen, Netherlands}
\emailAdd{federico.de.benedetti@cern.ch}

\abstract{In the new era of HL-LHC experiments, fast-timing detectors are emerging as critical tools for background rejection. Typical requirements include a temporal hit resolution of about \SI{50}{\pico\second}, a spatial resolution of around \SI{12}{\micro\meter}, and radiation hardness up to $10^{17}$~n$_\text{eq}$/cm$^2$. To address these challenges, the development of non-standard sensor designs and advanced fast readout electronics is required. The OPTIMA multichannel board addresses the need for testing small sensor demonstrators when they cannot yet be bonded to dedicated readout ASICs. It provides fast readout of up to 16 channels and can be integrated into various test setups, including test beam environments.
This contribution presents the design of the OPTIMA board, its integration in test beams, and the first experimental results.}


\begin{document}

\maketitle

\flushbottom
\section{Introduction}\label{sec:intro}
Detector upgrades for the High-Luminosity LHC and future experiments will face new challenges in terms of radiation tolerance, with integrated fluences up to $10^{17}$~n$_\text{eq}$/cm$^2$, while maintaining a high spatial resolution down to a few micrometres~\cite{ECFA}.
In the high occupancy conditions found, for example, in the inner tracking detectors, new approaches are needed to reduce the background due to the high pile-up. Introducing timing is the approach adopted by 4D tracking detectors such as the LHCb VELO in the second phase upgrade~\cite{U2TDR}. A time resolution of about \SI{50}{\pico\second} per hit is required to provide a \SI{20}{\pico\second} timestamp resolution per track. The associated R\&D must therefore identify suitable silicon sensor technologies capable of meeting these stringent timing and radiation-hardness requirements.

While the final characterisation focuses on hybridised sensor solutions, preliminary R\&D must be pursued using dedicated discrete fast front-end electronics capable of picosecond-level time resolution. The use of pixelated, small-pitch sensors introduces the need for a multichannel approach to study charge sharing between neighbouring pixels. To address these needs, the \ac{OPTIMA} board has been developed as a dedicated platform for the characterisation of non-hybridised fast silicon sensors.

\section{OPTIMA circuit}

\begin{figure}[htbp]
	\centering
	\begin{minipage}[c]{0.64\textwidth}
		\centering
		\includegraphics[width=\textwidth]{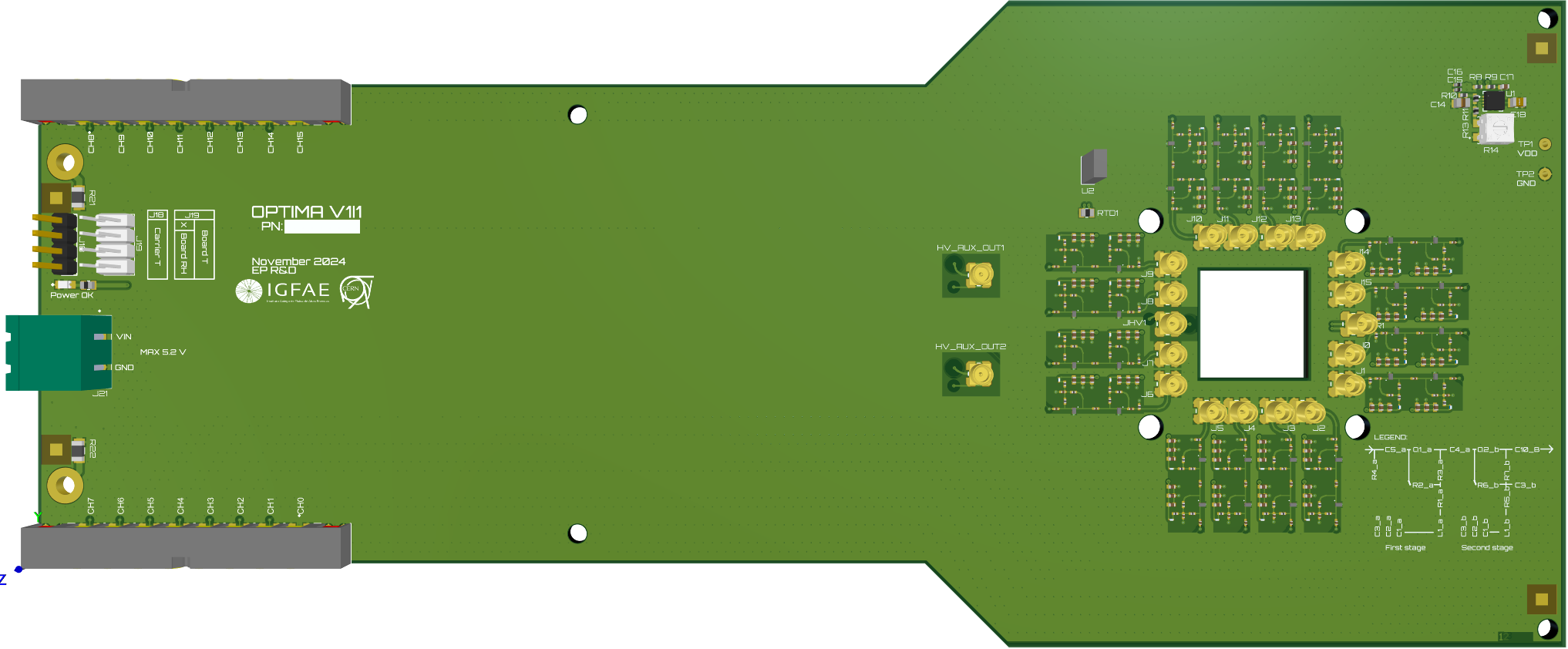}
	\end{minipage}%
	\qquad
	\begin{minipage}[c]{0.29\textwidth}
		\centering
		\includegraphics[width=\textwidth]{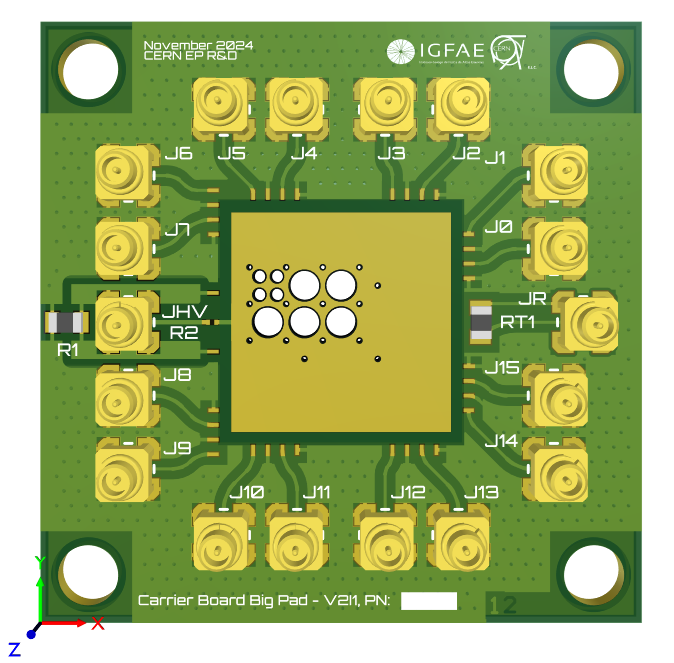}
	\end{minipage}
	\caption{\label{fig:optima_3d_cad} \ac{OPTIMA} motherboard (left) and carrier board (right).}
\end{figure}

The system consists of two main components, as shown in Figure~\ref{fig:optima_3d_cad}: a motherboard hosting the front-end electronics and a carrier board that holds the silicon sensors. This modular design allows sensor boards to be easily swapped during characterisation campaigns, irradiated independently, and wire-bonded according to the specific layout of different sensor prototypes.

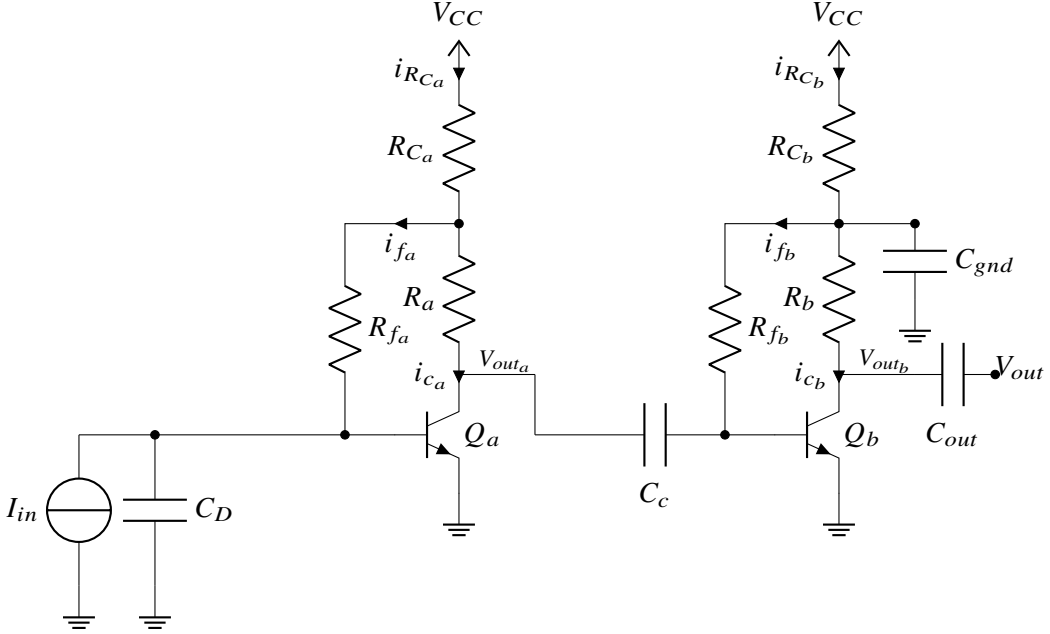
\begin{figure}[htbp]
	\centering
\def\NPNX{4}
\def\NPNY{3}
\def\RCOLBOT{\NPNY+0.8}
\def\RCOLTOP{\RCOLBOT+2}
\def\RONEBOT{\RCOLTOP}
\def\RONETOP{\RONEBOT+2}
\def\RFEEDX{\NPNX-1.5}
\def\IGENY{\NPNY-1}

\def\NPNXtwo{9}
\def\RTWOBOT{\RCOLTOP}
\def\RTWOTOP{\RTWOBOT+2}
\def\RFEEDXtwo{\NPNXtwo-1.5}

\begin{circuitikz}
    \draw(\NPNX,\NPNY) node[npn] (Q1) {$Q_a$};
    \draw(-1,\IGENY-1) to [I, l=$I_{in}$] (-1,\NPNY);      
    \draw(-1,\IGENY-1) node[ground] {};
    \draw(-1,\NPNY) -- (0,\NPNY);
    \draw(0,\NPNY-2) to [C, l_=$C_{D}$] (0,\NPNY) node[circ] {};
    \draw(0,\NPNY) -- (Q1.base); 
    \draw(0,\IGENY-1) node[ground] {};
    \draw(\NPNX,\RCOLBOT) node [label={[font=\footnotesize]right:$V_{out_a}$}, yshift=5pt] {};    
    \draw(Q1.collector) to [short, i<=${i}_{c_a}$] (\NPNX, \RCOLBOT) to                    
        [R, l=$R_a$] (\NPNX,\RCOLTOP) to
        [R, l=$R_{C_a}$] (\NPNX,\RONETOP) to
        [short, i<=${i}_{R_{C_a}}$] (\NPNX,\RONETOP) to
        node[vcc](vcc){$V_{CC}$} (\NPNX,\RONETOP);
    \draw(\NPNX, \RONEBOT) to [short, *-, i=${i}_{f_a}$ ] (\RFEEDX, \RONEBOT) to
        [R, l=$R_{f_a}$] (\RFEEDX,\NPNY) to
    (\RFEEDX,\NPNY) node[circ] {};
    \draw(Q1.emitter) node[ground] {};
    \draw(\NPNX, \RCOLBOT) -- (\NPNX+1, \RCOLBOT);
    \draw(\NPNX+1, \RCOLBOT) -- (\NPNX+1, \NPNY);
    \draw(\NPNXtwo,\NPNY) node[npn] (Q2) {$Q_b$};
    \draw(\NPNXtwo-4, \NPNY) to [C, l_=$C_{c}$] (Q2.base);
    \draw(\NPNXtwo,\RCOLBOT) node [label={[font=\footnotesize]right:$V_{out_b}$},yshift=5pt] {};
    \draw(Q2.collector) to [short, i<=${i}_{c_b}$] (\NPNXtwo, \RCOLBOT) to
        [R, l=$R_b$] (\NPNXtwo,\RCOLTOP) to
        [R, l=$R_{C_b}$] (\NPNXtwo,\RTWOTOP) to
        [short, i<=${i}_{R_{C_b}}$] (\NPNXtwo,\RTWOTOP) to
        node[vcc](vcc){$V_{CC}$} (\NPNXtwo,\RTWOTOP);
    \draw(\NPNXtwo, \RTWOBOT) to [short, *-, i=${i}_{f_b}$ ] (\RFEEDXtwo, \RTWOBOT) to
        [R, l=$R_{f_b}$] (\RFEEDXtwo,\NPNY) to
        (\RFEEDXtwo,\NPNY) node[circ] {};
    \draw(Q2.emitter) node[ground] {};
    \draw(\NPNXtwo+1,\RTWOBOT)  node[circ] {} to [C, l=$C_{gnd}$] (\NPNXtwo+1,\RCOLBOT+1) node[ground] {};
    \draw(\NPNXtwo+1,\RCOLBOT)  to [C, l_=$C_{out}$] (\NPNXtwo+2,\RCOLBOT) node[circ, right] {$V_{out}$};
    \draw(\NPNXtwo, \RTWOBOT)  -- (\NPNXtwo + 1, \RTWOBOT);
    \draw(\NPNXtwo, \RCOLBOT)  -- (\NPNXtwo + 1, \RCOLBOT);
\end{circuitikz}
	\caption{\label{fig:fe_circuit} Two-stage transimpedance amplifier circuit implemented in \ac{OPTIMA}.}
\end{figure}

The front-end circuit is based on a two-stage transimpedance amplifier derived from the USC single-channel board\footnote{\url{https://twiki.cern.ch/twiki/bin/view/Main/UcscSingleChannel\#UcscSingleChannel}} and is shown in Figure~\ref{fig:fe_circuit}. This circuit converts the sensor current impulse into a voltage signal suitable for readout with high-bandwidth oscilloscopes or fast digitisers such as the \ac{SAMPIC} acquisition system~\cite{SAMPIC}.

The core component is a heterojunction NPN SiGe BJT, selected for its high unity-gain bandwidth of \SI{75}{\giga\hertz}. A feedback resistor placed between the collector and the base nodes sets the transimpedance gain and reduces the input resistance seen by the sensor, making the circuit more efficient for current readout. The second amplification stage provides additional gain.  

The motherboard is equipped with a low-dropout voltage regulator, three independent high-voltage lines, and humidity and temperature sensors for testing under cold conditions. In this configuration, up to 16 channels can be read out simultaneously, and multiple structures can be tested with independent high-voltage and cooling capabilities. The onboard environmental sensors allow monitoring of the sensor temperature and ambient humidity, providing the necessary information to verify the dew point stability and the sensor temperature during cold operation.

The layout is implemented on a six-layer high-speed substrate (Rogers 4350B laminate), designed to provide a maximum bandwidth of \SI{6}{\giga\hertz}. A central \SI{18}{\times}\SI{18}{\milli\meter} cutout allows the particle beam to pass through the board. The compact component packaging allows the integration of multiple channels and minimises parasitic effects. With overall dimensions of \SI{80}{\times}\SI{255}{\milli\metre}, the \ac{OPTIMA} form factor allows the PCB to be used as a feed-through for setups where the sensors need to sit in a cold environment and/or in vacuum. It has been designed in particular to fit into the DUT boxes used for test-beam characterisation with the Timepix4 telescope~\cite{Telescope}.

The carrier board, shown in Figure~\ref{fig:optima_3d_cad} (right), is designed to host different sensor prototypes. It provides wire-bonding pads for up to 16 channels, includes a common bias-voltage line as well as a guard-ring connection, and features dedicated holes for laser injection. The carrier board is connected to the motherboard via coaxial connectors, ensuring signal integrity and minimising parasitic capacitance. Depth-controlled routing on the back of the board allows the thickness to be reduced in the central region down to \SI{200}{\micro\meter}, minimising the material budget along the beam line.

\section{Test beam results}\label{sec:test_beam_results}
A preliminary test-beam campaign was conducted in the summer of 2024 to validate the performance of \ac{OPTIMA}. The setup consisted of two \ac{MCP} detectors connected to an analog \ac{CFD} circuit, forming the time-reference system and achieving a time resolution of \SI{18}{\pico\second} with a single \ac{MCP}\cite{Telescope}.

\ac{OPTIMA} was installed in a dedicated enclosure mounted on a remotely controlled stage capable of translational and rotational motion. The acquisition system consisted of an MSO-B64B oscilloscope with a maximum bandwidth of \SI{10}{\giga\hertz} and a \ac{SAMPIC}~\cite{SAMPIC} module, the latter enabling multichannel acquisition with up to 32 channels recorded simultaneously.

The \ac{DUT}, shown in Figure~\ref{fig:three_stage_optima} on the right, was a 5 $\times$ 5 LGAD matrix sensor with a gain of approximately 20 and a target time resolution of \SI{30}{\pico\second}\cite{LGAD_CMS}, where 16 pixels were wire-bonded for multichannel readout. The sensor was operated at room temperature with a bias voltage ranging from \SI{180}{\volt} to \SI{230}{\volt}. The acquisition was triggered by the coincidence of the two \ac{MCP}s, and waveforms from both the oscilloscope and \ac{SAMPIC} were recorded for offline analysis. A \ac{CFD} algorithm was subsequently applied to the pixel signals to correct for the time-walk effect and extract the timing information. The time difference between the LGAD pixels and the \ac{MCP} reference was then computed to evaluate the time resolution.

\begin{figure}[htbp]
	\centering
	\begin{minipage}[c]{0.47\textwidth}
		\centering
		\includegraphics[width=\textwidth,trim=90 45 80 50,clip]{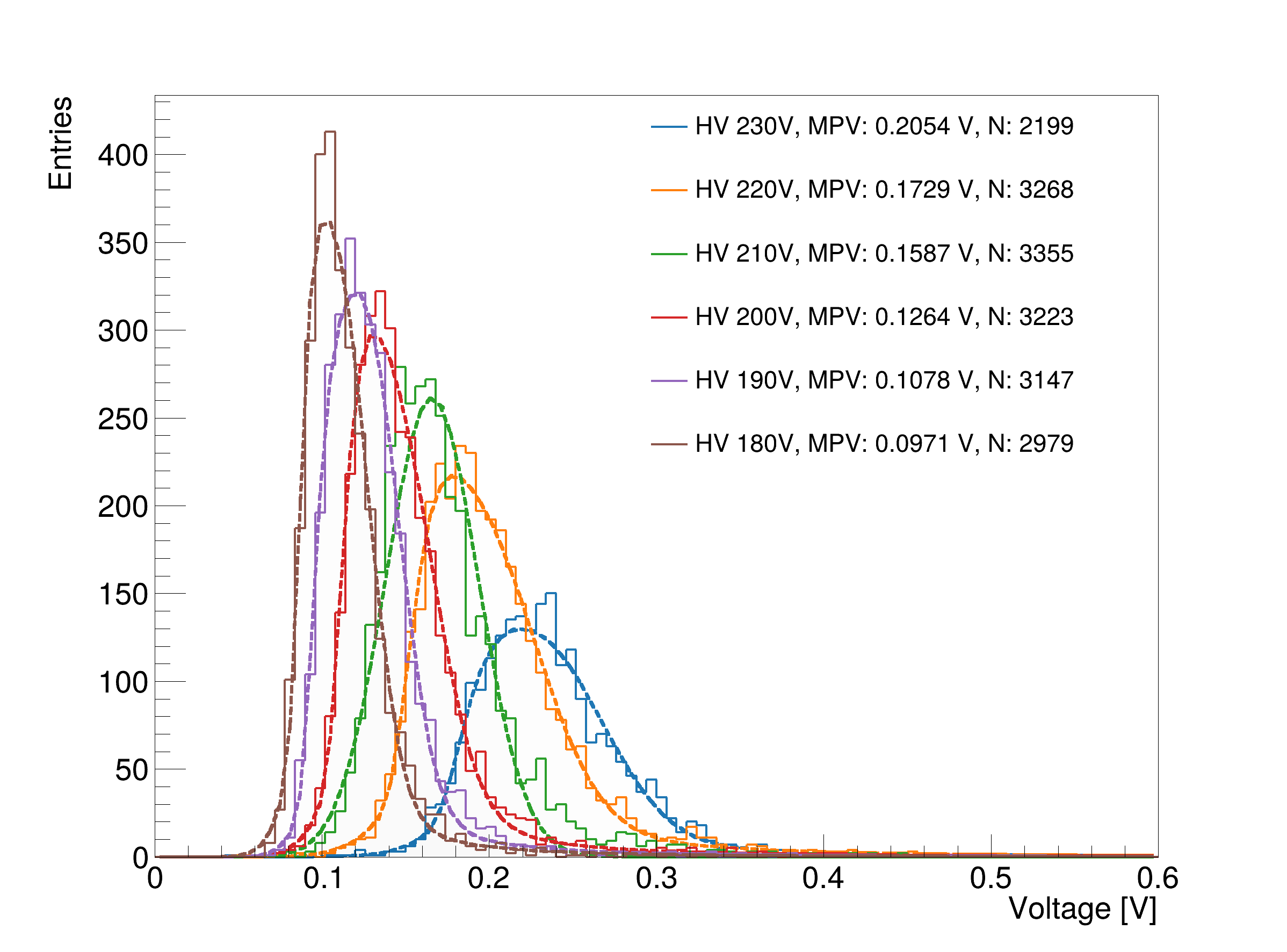}
	\end{minipage}%
	\qquad
	\begin{minipage}[c]{0.47\textwidth}
		\centering
		\includegraphics[width=\textwidth,trim=90 45 80 50,clip]{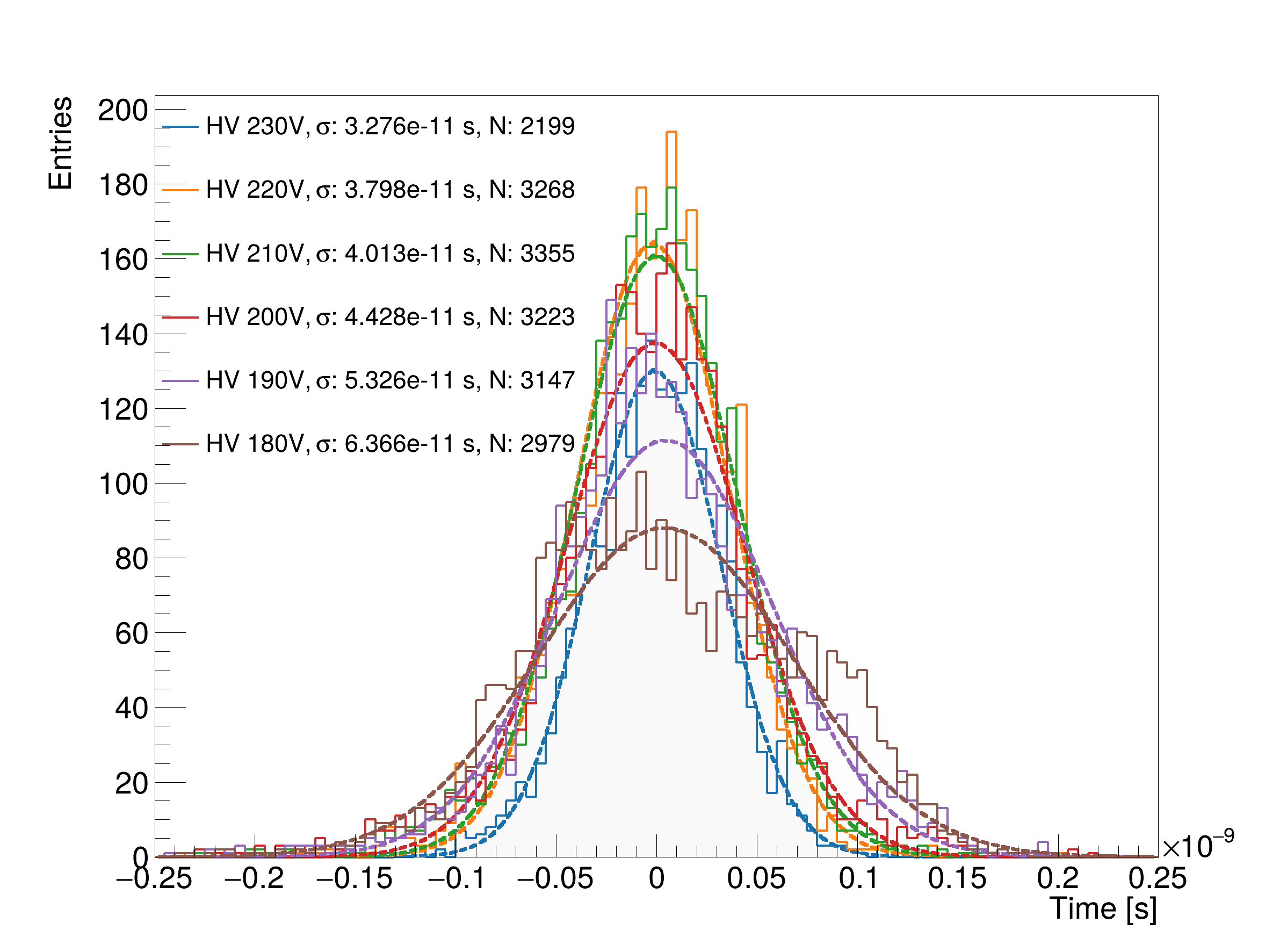}
	\end{minipage}
	\caption{\label{fig:lgad_time_resolution} Landau amplitude (left) and time-difference (right) distributions for the LGAD test-beam measurements. The time difference is computed between the LGAD pixel signal and the MCP reference. N denotes the number of entries in each distribution.}
\end{figure}

The results of the analysis are shown in Figure~\ref{fig:lgad_time_resolution}, where the Landau distribution (left) and the time-difference distribution (right) between the \ac{MCP} and the LGAD are displayed. At the highest applied bias voltage of \SI{230}{\volt}, the achievable time resolution is \SI{33}{\pico\second}, in good agreement with the expected value.

\section{Integration with Timepix4 telescope}\label{sec:timepix_integration}
\ac{OPTIMA} was designed for characterisation campaigns in conjunction with the Timepix4 telescope, capable of providing a pointing resolution of $\approx$ \SI{2.3}{\micro\meter} at the \ac{DUT} position in the centre~\cite{Telescope}. This characteristic allows a precise investigation of the \ac{DUT} performance using particle tracks to study charge collection, time response, and spatial uniformity at the pixel level.

\begin{figure}[htbp]
	\centering
	\begin{minipage}[c]{0.47\textwidth}
		\centering
		\includegraphics[width=\textwidth,trim=100 1100 100 1200,clip]{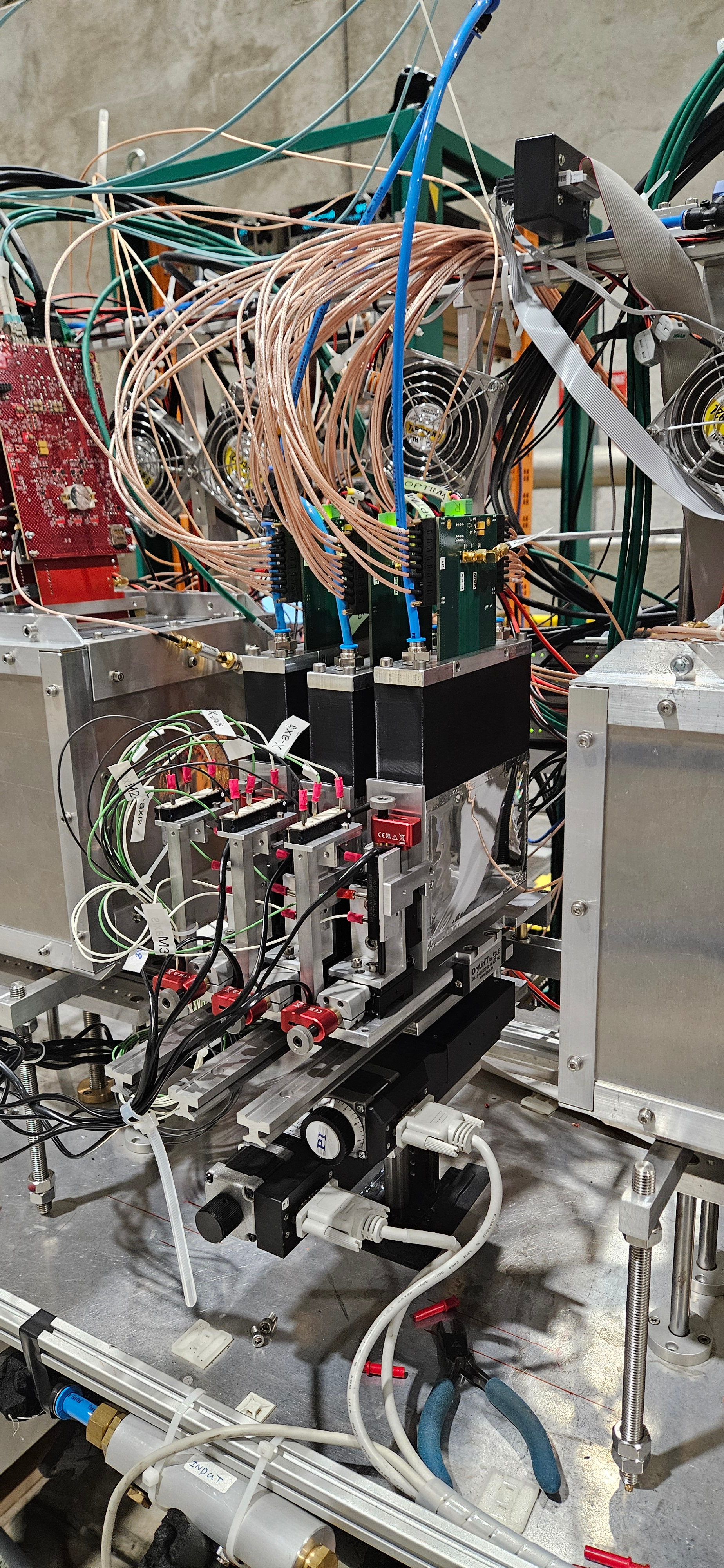}
	\end{minipage}%
	\qquad
	\begin{minipage}[c]{0.47\textwidth}
		\centering
		\includegraphics[width=\textwidth]{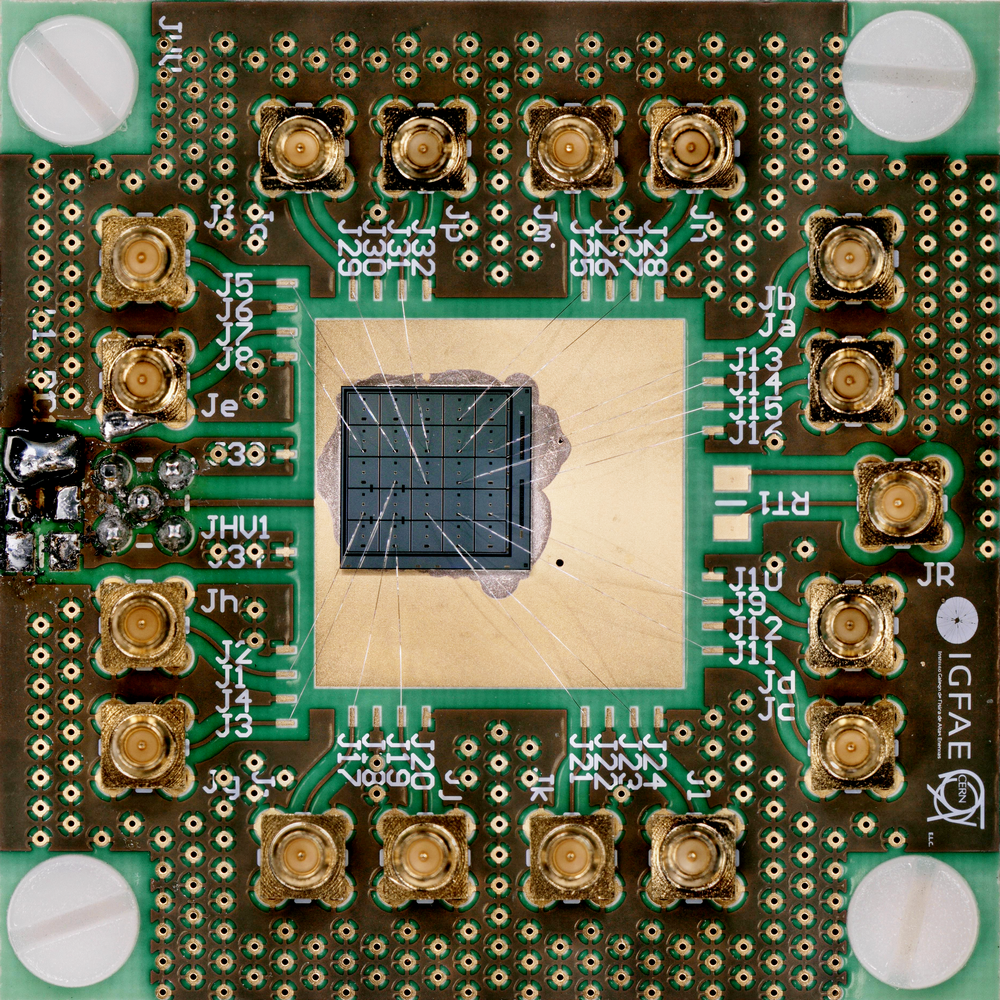}
	\end{minipage}
	\caption{\label{fig:three_stage_optima} Three-stage \ac{OPTIMA} setup installed in the centre of the Timepix4 telescope (left), LGAD matrix sensor with 16 wire-bonded channels (right).}
\end{figure}

The boards were installed into the middle stage of the Timepix4 setup, forming a three-stage configuration with three independent \ac{OPTIMA} as shown in Figure~\ref{fig:three_stage_optima} on the left. The X-Y translation stages can be adjusted separately, while the rotation stage is common to the entire structure. The mechanical setup was designed to maintain the beam centred on its barycentre during rotation.

The central \ac{OPTIMA} was equipped with the LGAD matrix sensor presented in Section~\ref{sec:test_beam_results} and used as a \ac{ROI} trigger. This configuration reduces the background on the \ac{DUT}s installed on the other \ac{OPTIMA}s, since it allows the use of a smaller pixel for the trigger rather than relying on the \ac{MCP}, which covers the full beam spot. The \ac{ROI} is configured as the central trigger in \ac{SAMPIC} to initiate the acquisition of the other \ac{DUT} channels, allowing efficiency measurements to be performed. The hits provided by \ac{SAMPIC} are timestamped and saved in a waveform-like data format for offline analysis.
\begin{figure}[htbp]
	\centering
	\begin{minipage}[c]{0.47\textwidth}
		\centering
		\includegraphics[width=\textwidth,trim = 0 0 0 50,clip]{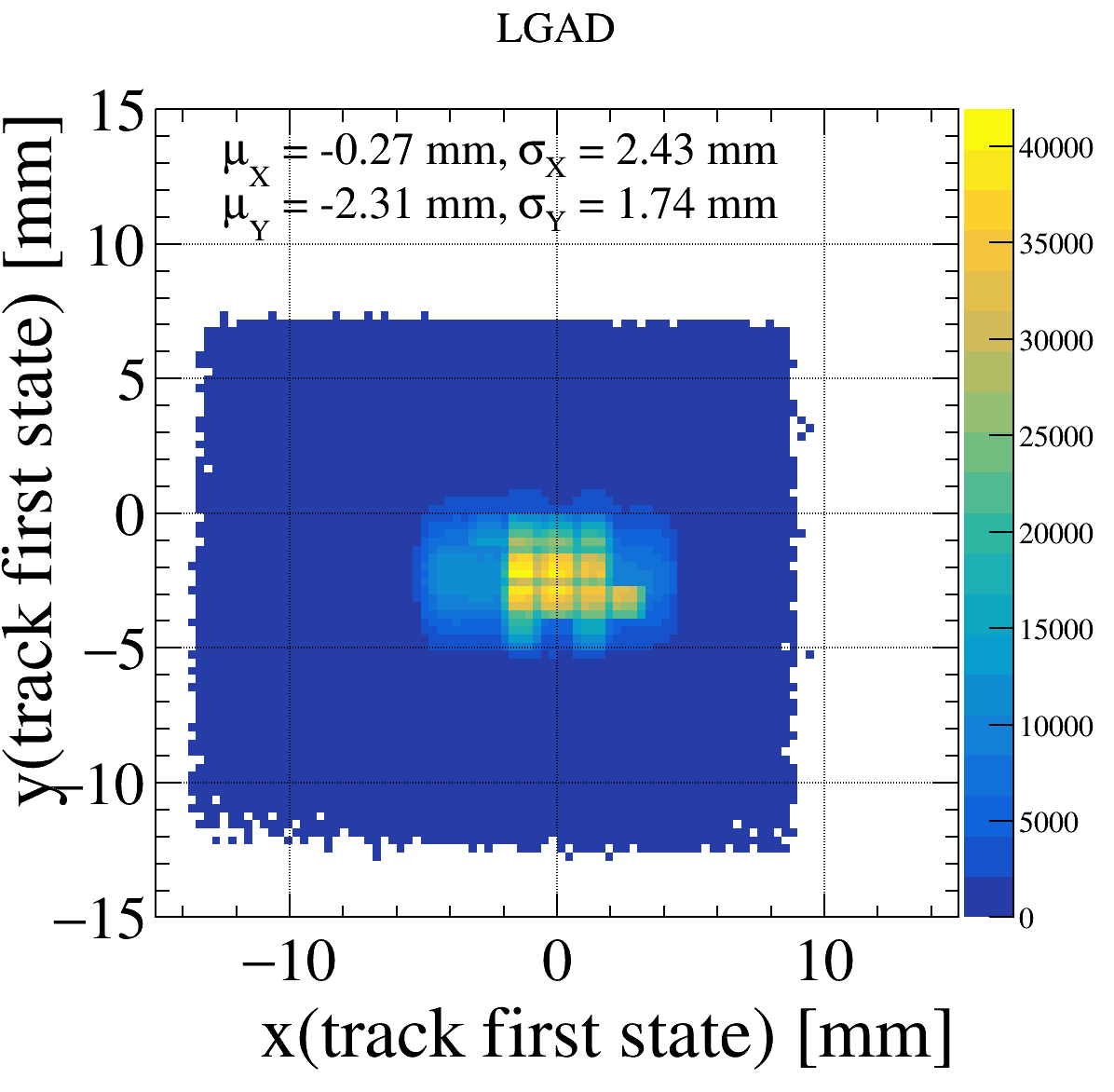}
	\end{minipage}%
	\qquad
	\begin{minipage}[c]{0.47\textwidth}
		\centering
		\includegraphics[width=\textwidth,trim = 0 0 0 50,clip]{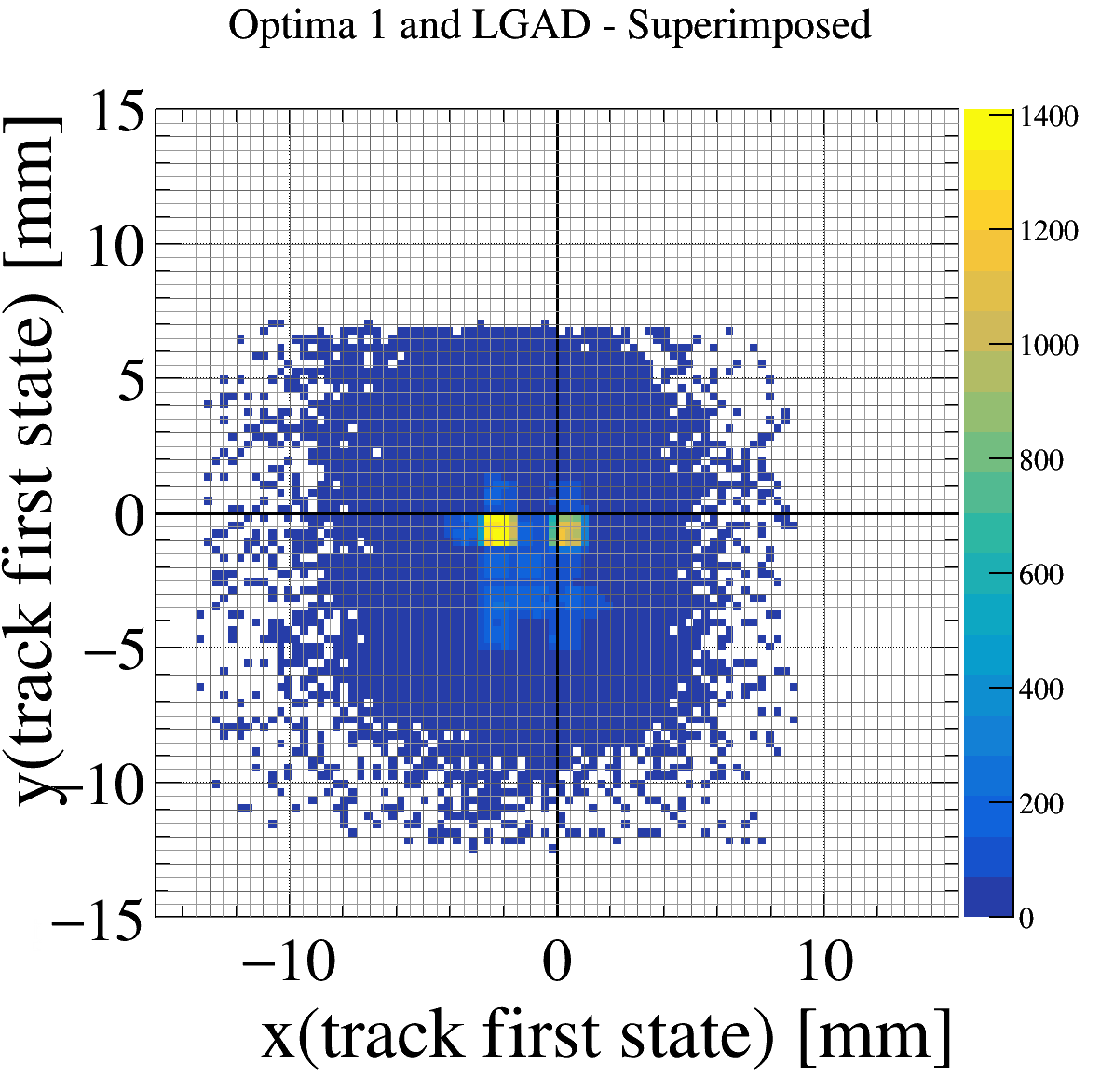}
	\end{minipage}
	\caption{\label{fig:telescope_space_alignemnt} Hit maps obtained from SAMPIC after alignment with Timepix4 telescope tracks: LGAD Region of Interest (ROI) used as trigger (left) and corresponding beam projection on the Device Under Test (DUT) (right).}
\end{figure}
To maintain temporal synchronisation with the telescope, a 100 MHz clock was derived from the main Timepix4 clock, while a T0 synchronisation signal provides the start-of-run command to the entire system. A dedicated data decoder was implemented in Kepler\footnote{https://gitlab.cern.ch/TimepixTelescope/Kepler/}, the telescope analysis framework. A time-correlation algorithm was developed to associate telescope tracks with \ac{SAMPIC} hits. The resulting spatial alignment with the beam is shown in Figure~\ref{fig:telescope_space_alignemnt}, where the \ac{ROI}-to-beam projection and the \ac{DUT} positions can be observed after the track association is performed.

The successful integration of the \ac{OPTIMA} system within the Timepix4 telescope setup enables the correlation of the timing information provided by \ac{SAMPIC} with the track position reconstructed by the Timepix4 telescope across multiple pixels, allowing a precise spatial and temporal characterisation of the device under test.

\section{Conclusions}
\ac{OPTIMA} has been developed as a versatile and modular platform for the characterisation of non-hybridised silicon sensors. The system integrates high-bandwidth front-end electronics, environmental monitoring, and flexible biasing and cooling capabilities, making it suitable for a wide range of sensor technologies and operating conditions.

The preliminary test-beam results confirm the excellent performance of the readout chain, achieving a time resolution consistent with the expected values for gain-layer sensors, with a target resolution below \SI{50}{\pico\second}. The system demonstrates stable operation and signal integrity across all 16 readout channels, validating its capability for multichannel sensor studies.

The successful integration of \ac{OPTIMA} within the Timepix4 telescope infrastructure represents an important step toward a complete characterisation of non-hybridised silicon detectors both in space and time. This setup enables precise correlation between the timing information provided by \ac{SAMPIC} and the spatial track reconstruction from the telescope, allowing detailed studies of inter-pixel behaviour, efficiency, and timing uniformity across the sensor area.

Ongoing developments focus on optimising the front-end electronics for non-gain silicon sensors, aiming to extend the platform’s applicability to a broader class of detector technologies. Future test-beam campaigns are planned to validate these improvements and to perform systematic studies on irradiated sensors.

\acknowledgments%
This work was carried out in the context of the CERN Strategic R\&D Programme on Technologies for Future Experiment https://ep-rnd.web.cern.ch/.

\bibliographystyle{JHEP}
\bibliography{references}

\end{document}